\documentclass[aps,amsmath,amsfonts,11pt]{revtex4}
\usepackage{graphicx}
\usepackage{epsfig,graphicx}
\usepackage[english]{babel}
\usepackage{amsfonts}
\usepackage{amsmath}
\usepackage{latexsym}
\usepackage{graphics,bm}
\usepackage{dcolumn}
\usepackage{bm}
\usepackage{rotating}

\begin{document}

\title{Spin and tunneling dynamics in an asymmetrical double quantum dot with spin - orbit coupling}

\author{Madhav Singh$^{1}$,  Pradeep K Jha$^{2}$ and  Aranya B Bhattacherjee$^{3}$}

\address{$^{1}$ Department of Physics and Astrophysics, University of Delhi, Delhi-110007, India \\
$^{2}$ Department of Physics, DDU College, University of Delhi, New Delhi \\
$^{3}$ School of Physical Sciences, Jawaharlal Nehru University, New Delhi-110067, India }

\begin{abstract}

In this article we study the spin and tunneling dynamics as a function of magnetic field in a one-dimensional GaAs double quantum dot with  both the Dresselhaus and Rashba spin-orbit coupling.  In particular we consider different spatial widths for the spin-up and spin-down electronic states. We find that the spin dynamics is a superposition of slow as well as fast Rabi oscillations. It is found that the Rashba interaction strength as well as the external magnetic field strongly modifies the slow Rabi oscillations which is particularly useful for single qubit manipulation for possible quantum computer applications.

\end{abstract}

\maketitle
\section{Introduction}

Semiconductor quantum dots (QDs), also called ``artificial atoms'' are analogous to real atoms \citep{bird} and have been one of the most intensively studied nanostructures due to the richness in their properties \citep{kohler}. Numerous interesting transport properties of QD based systems have been revealed such as Coulomb blockade, spin blockade and Kondo effect \citep{recher, weinmann, takahashi, liang, park}. One of the challenges in spintronics and quantum transport is fast and efficient manipulation of electron spins in QDs.  Infact by controlling the electric bias, tunable spin diodes have been realized experimentally \citep{merchant, hamaya}. Manipulated electron spins in QDs are expected as a possible realization of qubits \citep{burkard}. In the presence of spin-orbit (SO) coupling, single electron QDs exhibit spin-flip dynamics which can be understood in terms of the electric dipole spin resonance (EDSR) \citep{rashba1, rashba2}. Using EDSR, coherent spin manipulation in GaAs QDs have been demonstrated, where electric field induced spin Rabi oscillations have been observed \citep{nowack}. Other various SO effects present in semiconductor nanostructures provide an efficient and reliable way to manipulate electron spins in two-dimensional (2D) electron gases \citep{winkler, pershinnd, sheng, rasanen, kuan, liu, joibari, gumber}. Manipulating spin degrees of freedom has opened new possibilities to design fast and low power quantum devices for applications in quantum computing and memory storage \citep{wolf, awschalom}.

However for quantum information applications, single isolated QDs are not suitable since interdot interaction is necessary to produce and manipulate many-body states. Double QDs where tunneling plays a crucial role are more promising for quantum information technologies \citep{petta}. The interplay between tunneling and spin flip process is an important factor in double QDs. In the presence of SO interaction, the electron states and the interdot tunneling becomes spin-dependent.

Single electron dynamics in a one-dimensional double QD with SO coupling driven by an external electric and magnetic fields has been a subject of earlier numerous study \citep{sherman1, sherman2, sherman3, sherman4}. In these works, the mutual effect of coordinate and spin motion on the Rabi oscillations were carried out under the assumption that the spatial widths of the wacefunction for spin-up and spin-down states are same along the direction of the confining potential. However since the magnetic field lifts the spin degeneracy, the wavefunctions in the two spin-split levels are not the same \citep{bhowmik}. Keeping this important point in view, in this article we study the spin and tunneling dynamics as a function of magnetic field in a one-dimensional double QD with SO coupling and different spatial widths for the spin-up and spin-down states. Both the Dresselhaus and Rashba SO coupling are taken into account.

\section{The Model and Hamiltonian}

We consider an electron confined in a one dimensional double QD described by the quadratic potential as shown in figure1.

\begin{eqnarray}\label{eqn1}
V(x)= U_0(\frac{-2x^2}{d^2}+\frac{x^4}{d^4}),
\end{eqnarray}

where the two minima located at d and -d are separated by a barrier of height $U_0$.

\begin{figure}[t]
\vspace{0.2cm}
\hspace{0.0cm}
\centering
\mbox{\includegraphics[scale=0.7]{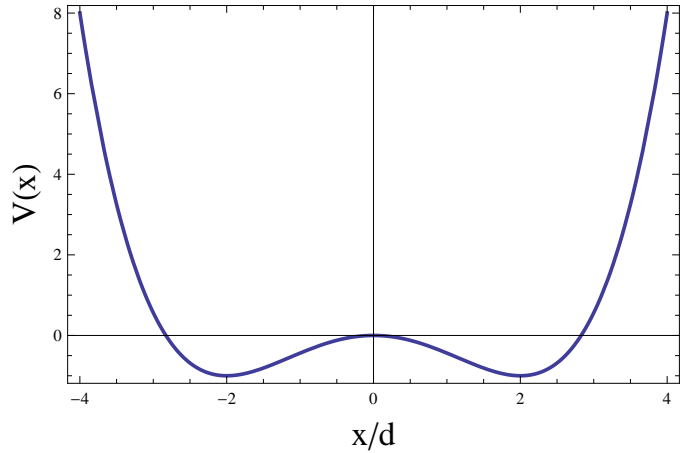}}
\caption{Schematic plot of the double well potential described by Equation (1). }
\end{figure}\label{fig1}

The unperturbed system is described by the Hamiltonian:

\begin{equation}\label{eqn2}
H_0=\frac{\hat{p}_x^2}{2m}+{V(\hat{x})} ,
\end{equation}

where $\hat{p}_x = -i \hbar \frac{\partial}{\partial x}$ is the momentum operator in the $x-$ direction. Here m is the mass of the electron. In the absence of any external fields and spin-orbit coupling (SOC), the ground state is split into the doublet of even (symmetric) and odd (anti symmetric) states. The tunneling energy $\Delta E_g << V_0 $ determines the tunneling time $ T_{tun} = \frac{2 \pi}{\Delta E_g}$ and $\Delta E_g$ also is the gap between the ground and first excited state of $H_0$. Under the influence of an external magnetic field $B_z$ along $Z$ direction and electric field along the x direction the system is described by Hamiltonian

\begin{equation}\label{eqn3}
H=H_0-eE x+H_{SO}-\frac{\Delta_z \sigma_z}{2}.
\end{equation}

The second term $-eEx$ describe the perturbation caused by external electric field. The Zeeman coupling to the magnetic field is $\frac{\Delta_{z}\sigma_z}{2}$, where $\Delta_z=\mid g\mid \mu_{B} B_{z} $ is the Zeeman splitting. The Pauli matrices are described by  $\sigma_{i=x,y,z}$. Here $g$ is the Land\'{e} factor of electron and $\mu_B$ is the Bohr magneton. The spin –orbit (SO) interaction is described by Hamiltonian $H_{SO}$ which is the sum of the bulk –originated Dresselhaus$(\beta)$ and structure – related Rashba $(\alpha )$ terms,

\begin{equation}\label{4}
H_{SO}= (\frac{\beta}{\hbar}p_x \sigma_x- \frac{\alpha}{\hbar}p_x \sigma_y ).
\end{equation}

In order to study the spin and tunnelling dynamics as a function of the applied magnetic field , we use a perturbation approach and diagonalize the Hamiltonian $H$ in the truncated spinor basis $\Psi_{m}(x)|\uparrow>$  with corresponding eigenvalues $E_{m,\sigma}$. Here  $\Psi_{m}(x)$ are the eigenfunctions of $H_{0}$ in the double well potential with $ m=s$ (symmetric), $m=a$ (antisymmetric) and $\sigma= +(-)$ corresponds to the spin parallel (antiparallel) to the $z$ axis. The wavefunction$\Psi_{s}(x)$($\Psi_{a}(x)$) is even (odd) with respect to the inversion of x. Assuming weak tunneling, these symmetric and antisymmetric functions can be written in the form :

\begin{equation}\label{5}
\Psi_{s,a}(x)=\frac{\Psi_L(x)\pm \Psi_R(x)}{\sqrt{2}},
\end{equation}

where $\Psi_L(x)$ and $\Psi_R(x)$ are localized wavefunction in the left and right dot repectively.

\begin{equation}\label{6}
\Psi_L(x)=\frac{1}{\sqrt{8 w_x}} \sin(\frac{\pi}{2}+\frac{nx\pi}{w_x}),
\end{equation}

\begin{equation}
\Psi_R(x)=\frac{1}{\sqrt{8 w_x}} \sin(\frac{\pi}{2}-\frac{nx\pi}{w_x}).
\end{equation}

Here $w_x$ is the spatial width of the wavefunction. We will be restricting our analysis to two lowest states $n = 1,2 $. The first four lowest energy states are $|\Psi_1>=\Psi_s(x)|\uparrow>$,$|\Psi_2>=\Psi_s(x)|\downarrow>$,$|\Psi_3>=\Psi_a(x)|\uparrow>$ and $|\Psi_4>=\Psi_a(x)|\downarrow>$. The spatial parts of the wavefunctions $\Psi_{s(a)}(x)$ are different for spin-up and spin-down states. The spin degeneracy is lifted due to the magnetic field and as a result of which the wavefunctions in the two spin split levels are not the same. Hence the spatial spread of the wavefunctions are different . The wavefunction corresponding to the state with higher energy spreads out more outside the potential well (spatial width $w_{x} ^{'}$ ) compared to the wavefunction corresponding to the state width lower energy level (spatial width $w_x$ ) and $w_{x} ^{'} > w_{x}$. The full dynamics of the system can then be studied with the function

\begin{figure}[t]
\hspace{-0.0cm}
\begin{tabular}{cc}
\includegraphics [scale=0.60]{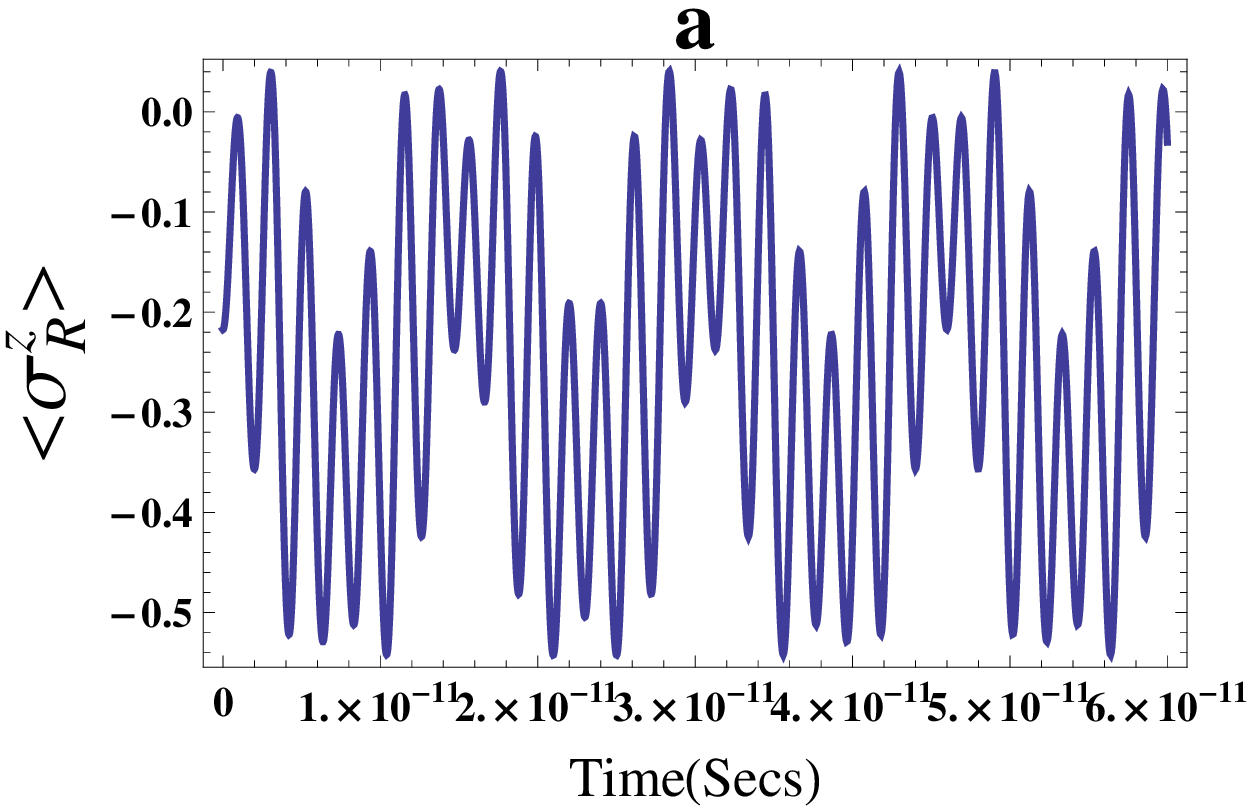}& \includegraphics [scale=0.60]{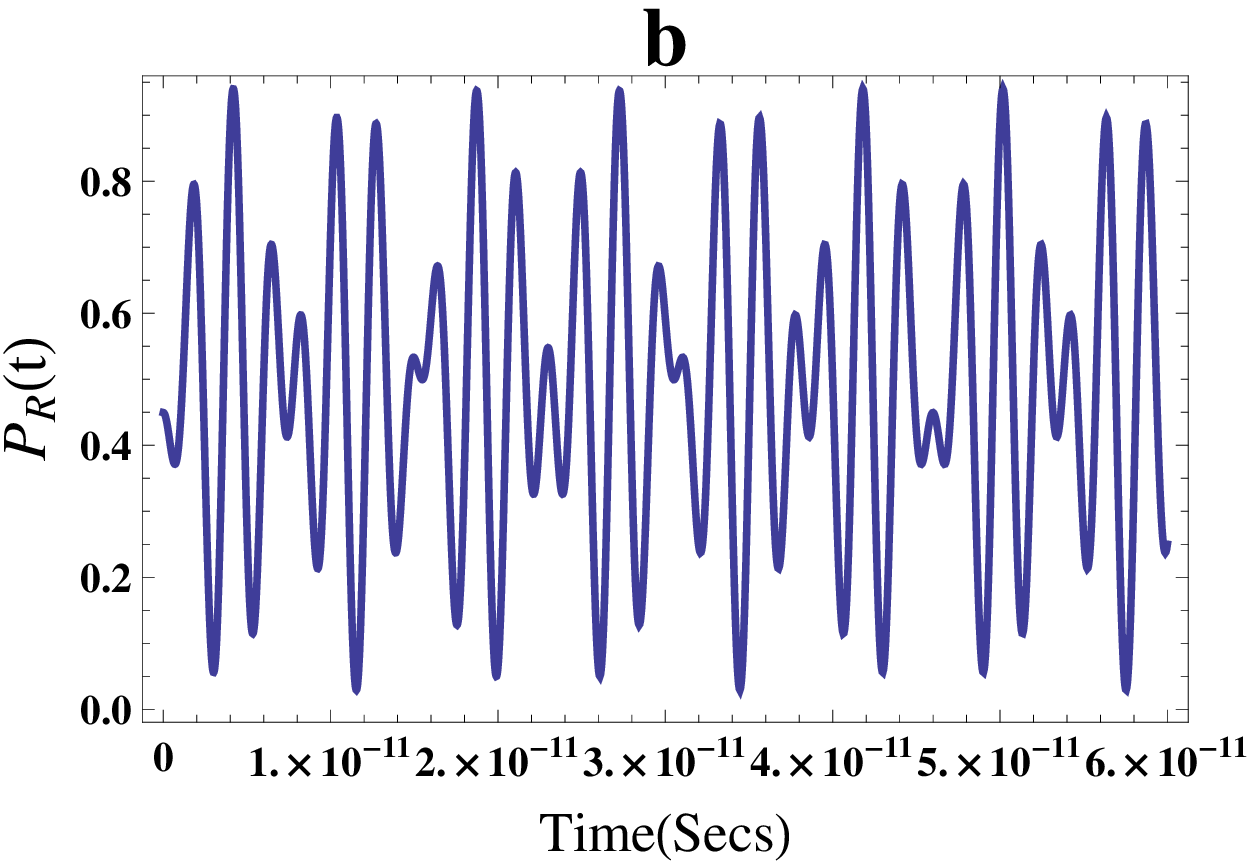}\\
\end{tabular}
\caption{Dynamics of $\sigma_{R}^{z}(t)$ (a) and $P_{R}(t)$ (b) for the identical width (ID) case for $B=0.88 T$. The various parameters used are: $\alpha=1 \times 10^{-9} eV-cm$, $\beta=0.3 \times eV-cm$, g=-0.45, $w_{x}=25 \sqrt{2} nm$, $w_{x}=w_{x}^{'}$ }
\end{figure}\label{fig2}

\begin{figure}[t]
\hspace{-0.0cm}
\begin{tabular}{cc}
\includegraphics [scale=0.60]{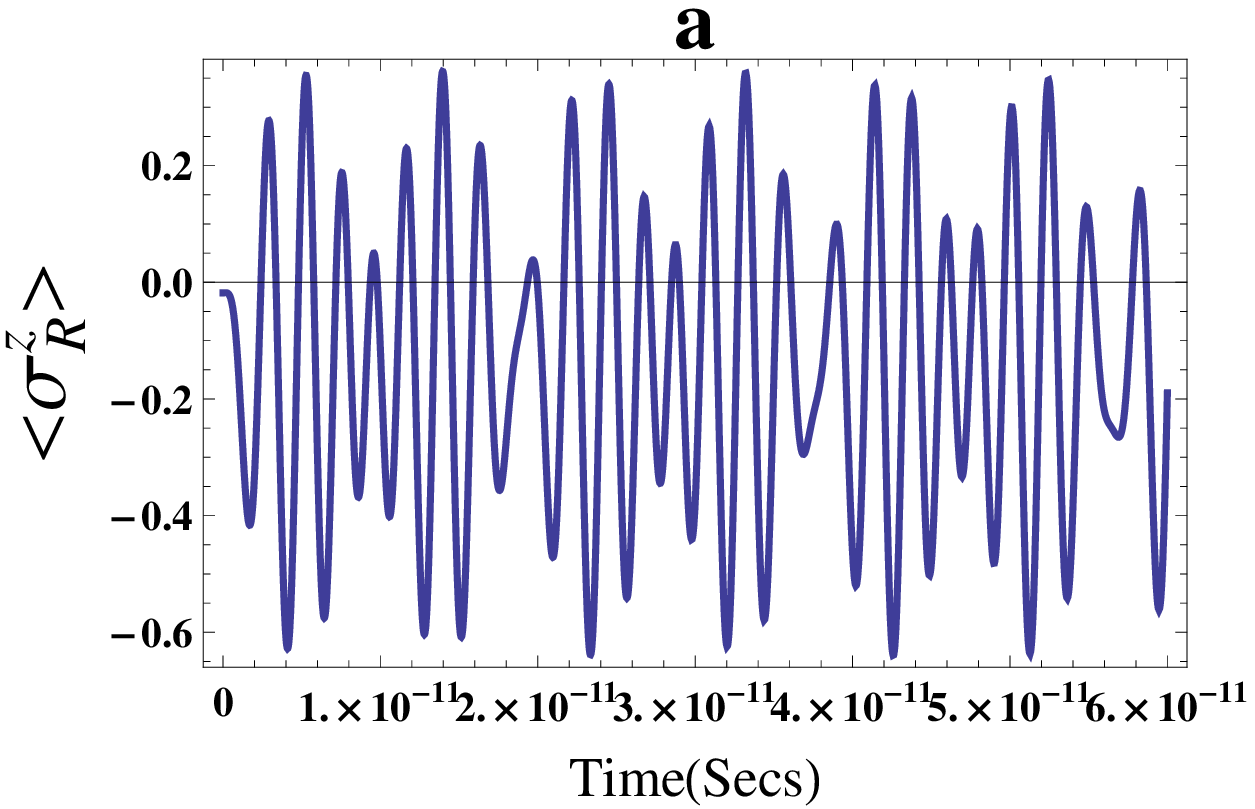}& \includegraphics [scale=0.60]{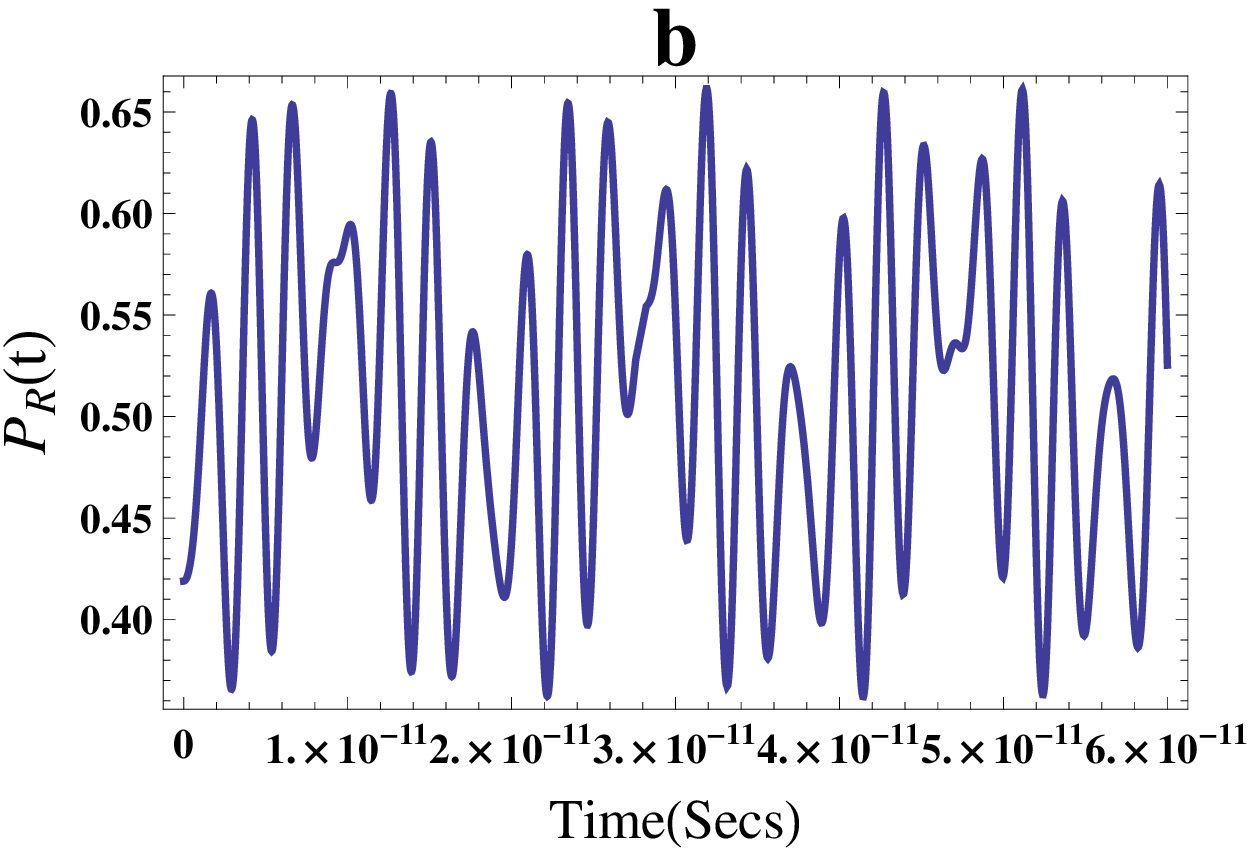}\\
\end{tabular}
\caption{Dynamics of $\sigma_{R}^{z}(t)$ (a) and $P_{R}(t)$ (b) for the identical width (ID) case for $B=6.4 T$. The various parameters used are: $\alpha=1 \times 10^{-9} eV-cm$, $\beta=0.3 \times eV-cm$, g=-0.45, $w_{x}=25 \sqrt{2} nm$, $w_{x}=w_{x}^{'}$ .  }
\end{figure}\label{fig3}

\begin{figure}[t]
\hspace{-0.0cm}
\begin{tabular}{cc}
\includegraphics [scale=0.60]{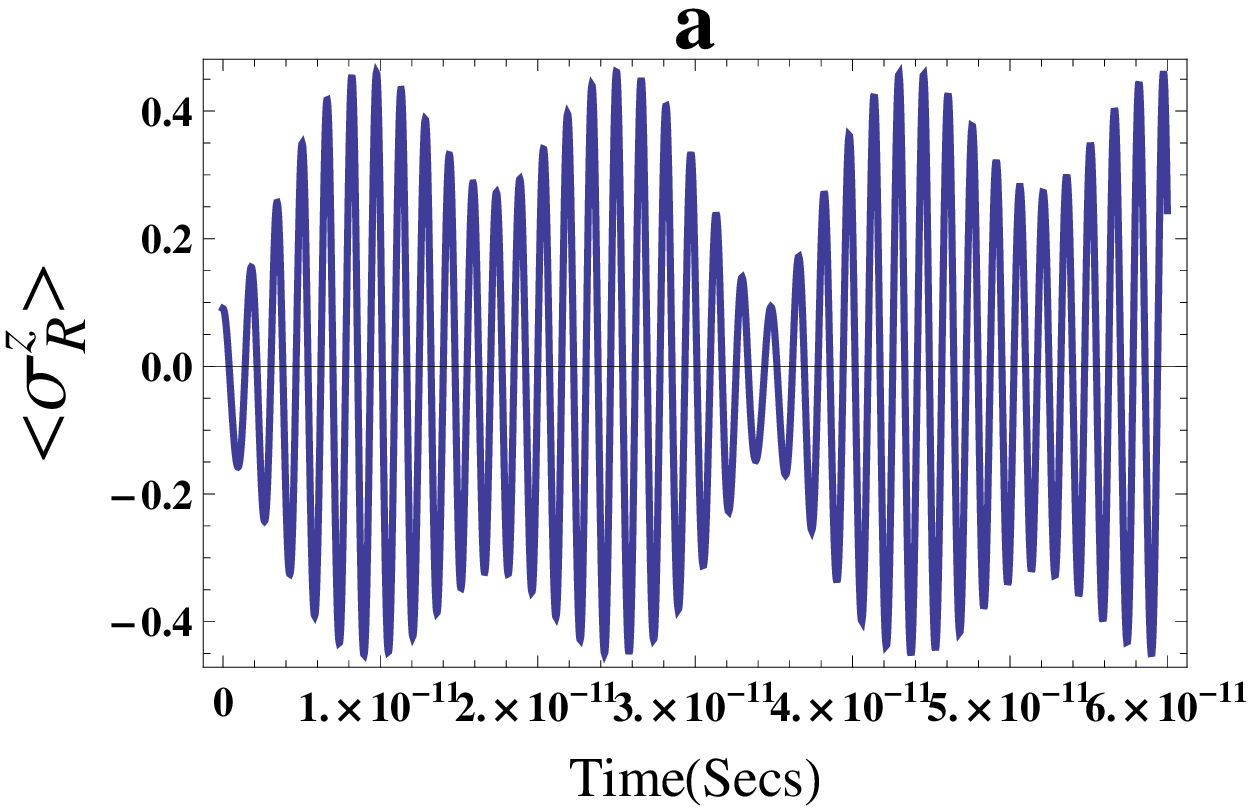}& \includegraphics [scale=0.60]{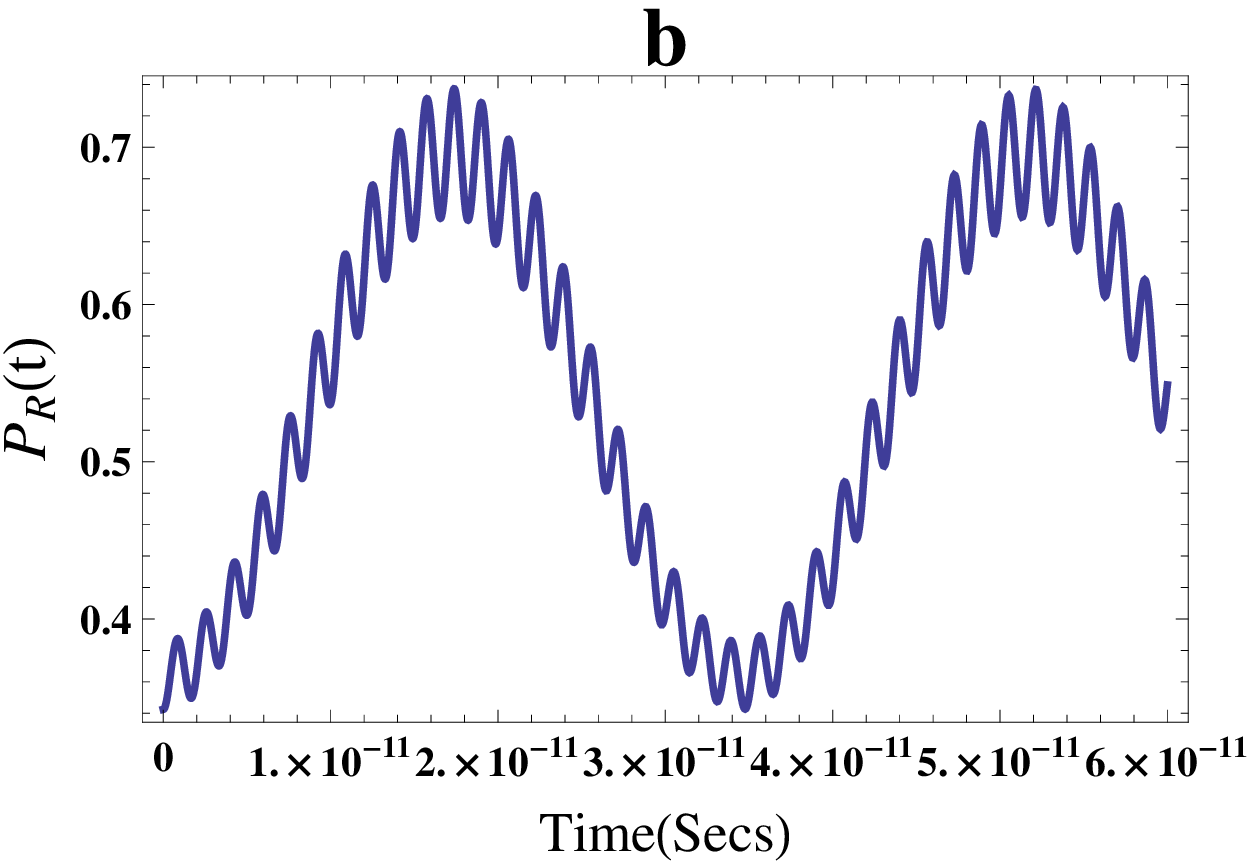}\\
\end{tabular}
\caption{Dynamics of $\sigma_{R}^{z}(t)$ (a) and $P_{R}(t)$ (b) for the non-identical width (NID) case for $B=0.88 T$. The various parameters used are: $\alpha=1 \times 10^{-9} eV-cm$, $\beta=0.3 \times eV-cm$, g=-0.45, $w_{x}=25 \sqrt{2} nm$, $w_{x}^{'}=1.0375 w_{x}$ .  }
\end{figure}\label{fig4}

\begin{figure}[t]
\hspace{-0.0cm}
\begin{tabular}{cc}
\includegraphics [scale=0.60]{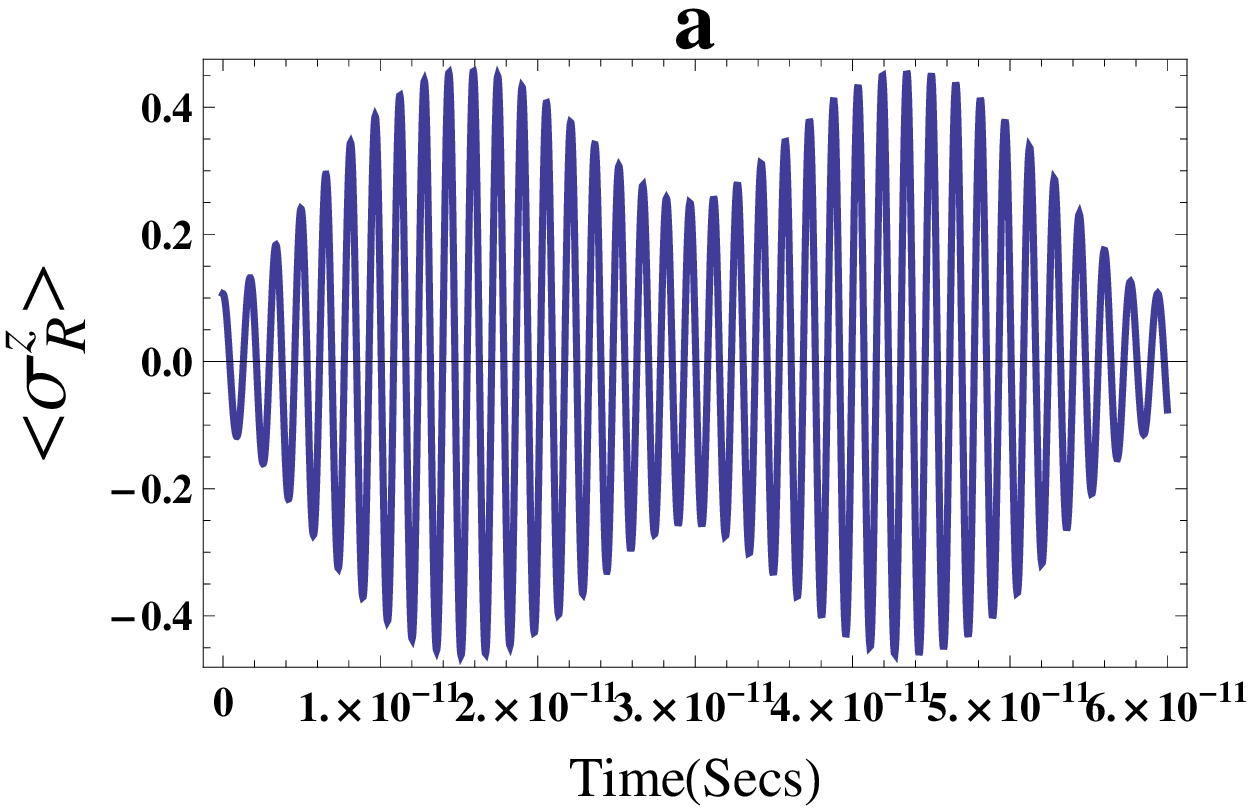}& \includegraphics [scale=0.60]{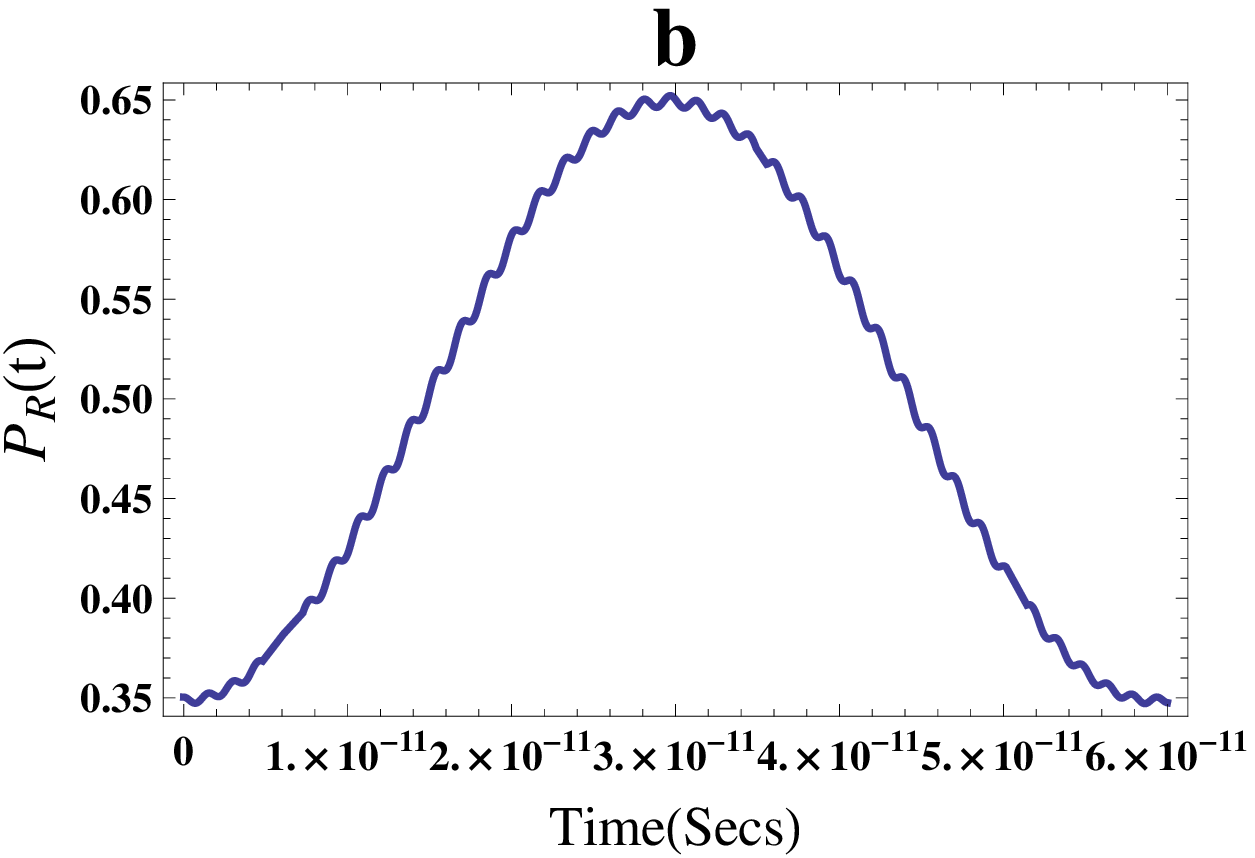}\\
\end{tabular}
\caption{Dynamics of $\sigma_{R}^{z}(t)$ (a) and $P_{R}(t)$ (b) for the non-identical width (NID) case for $B=6.4 T$. The various parameters used are: $\alpha=1 \times 10^{-9} eV-cm$, $\beta=0.3 \times eV-cm$, g=-0.45, $w_{x}=25 \sqrt{2} nm$, $w_{x}^{'}=1.0375 w_{x}$ .  }
\end{figure}\label{fig5}

\begin{figure}[t]
\hspace{-0.0cm}
\begin{tabular}{cc}
\includegraphics [scale=0.60]{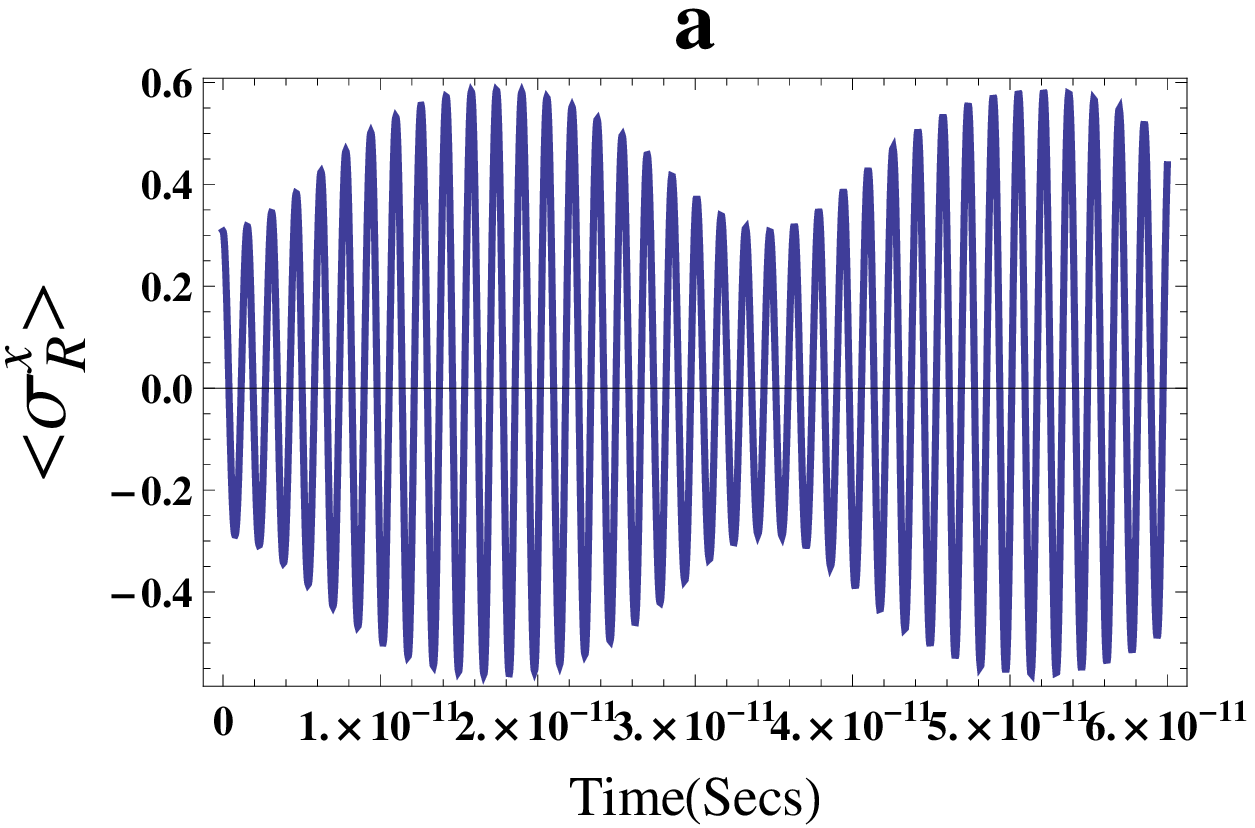}& \includegraphics [scale=0.60]{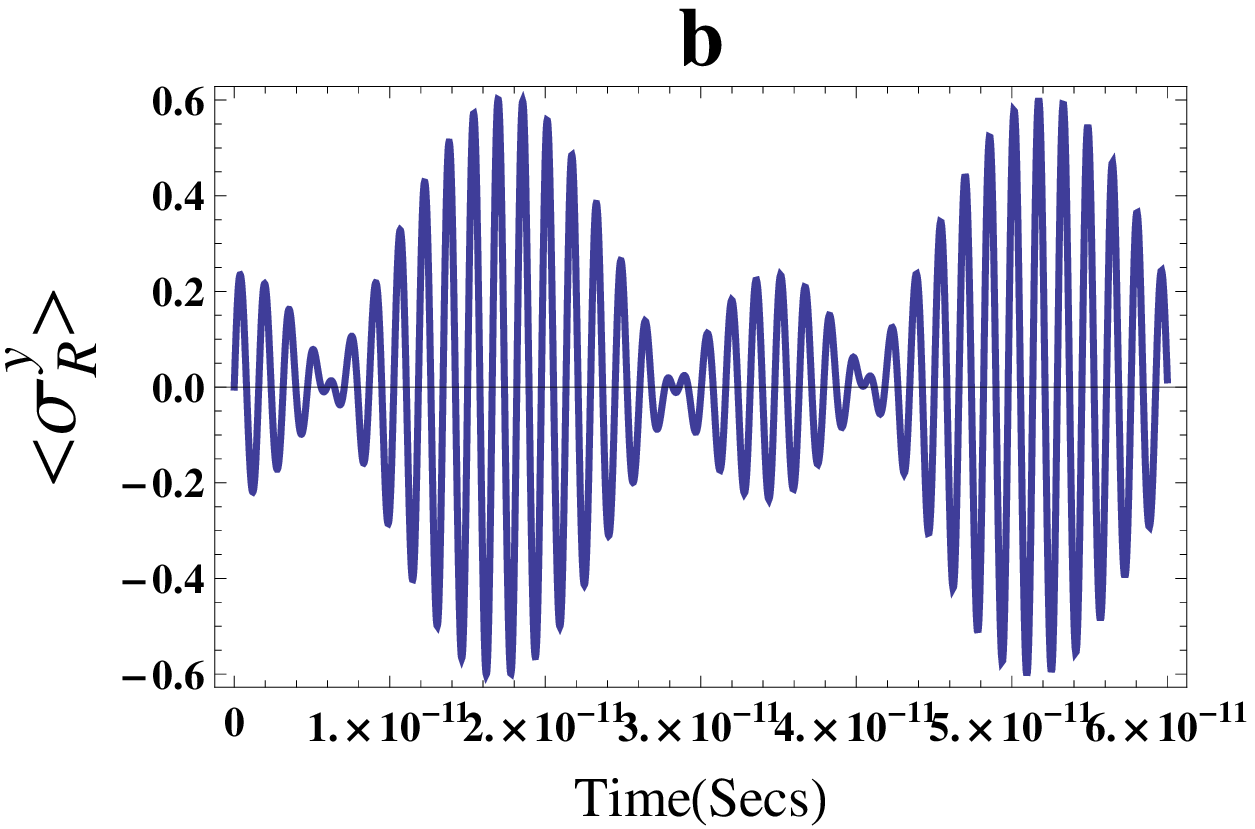}\\
\end{tabular}
\caption{Dynamics of $\sigma_{R}^{x}(t)$ (a) and  $\sigma_{R}^{y}(t)$ (b) for the non-identical width (NID) case for $B=0.88 T$. The various parameters used are: $\alpha=1 \times 10^{-9} eV-cm$, $\beta=0.3 \times eV-cm$, g=-0.45, $w_{x}=25 \sqrt{2} nm$, $w_{x}^{'}=1.0375 w_{x}$. }
\end{figure}\label{fig6}

\begin{figure}[t]
\hspace{-0.0cm}
\begin{tabular}{cc}
\includegraphics [scale=0.60]{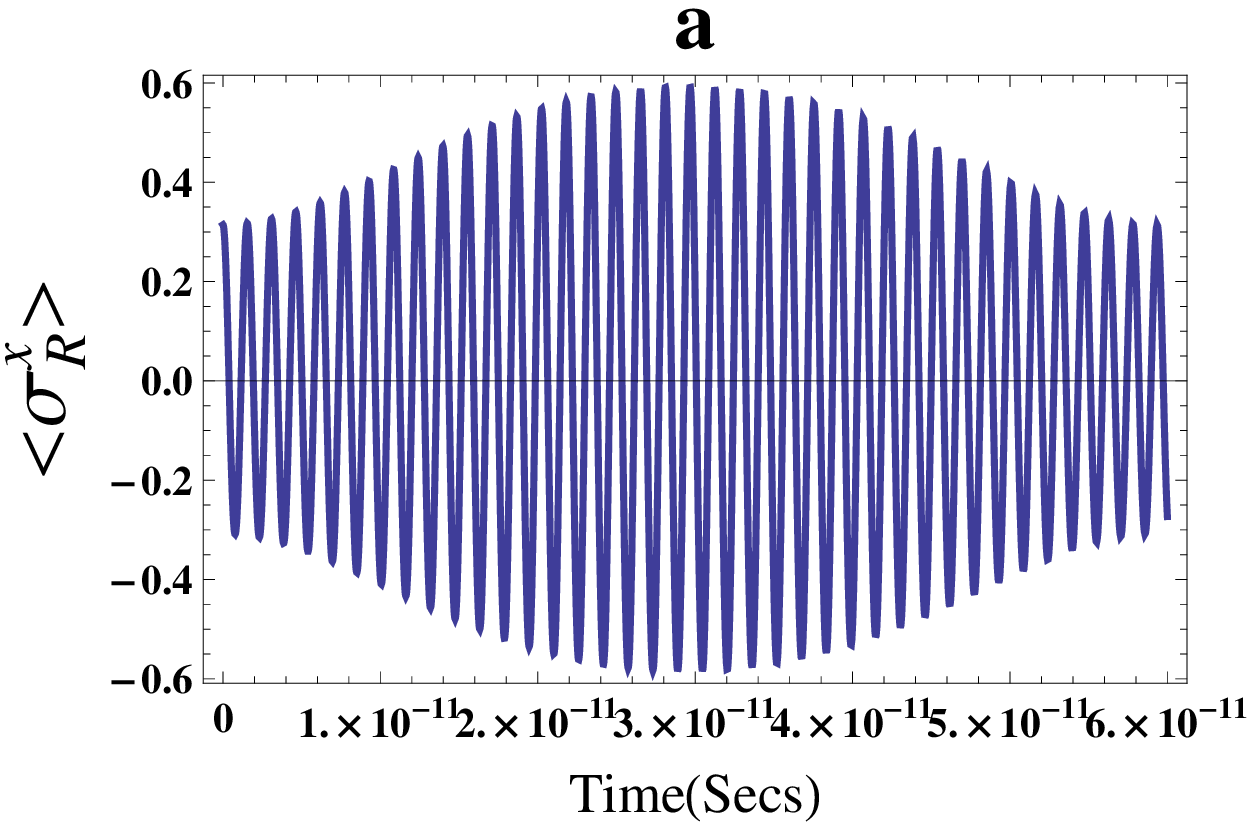}& \includegraphics [scale=0.60]{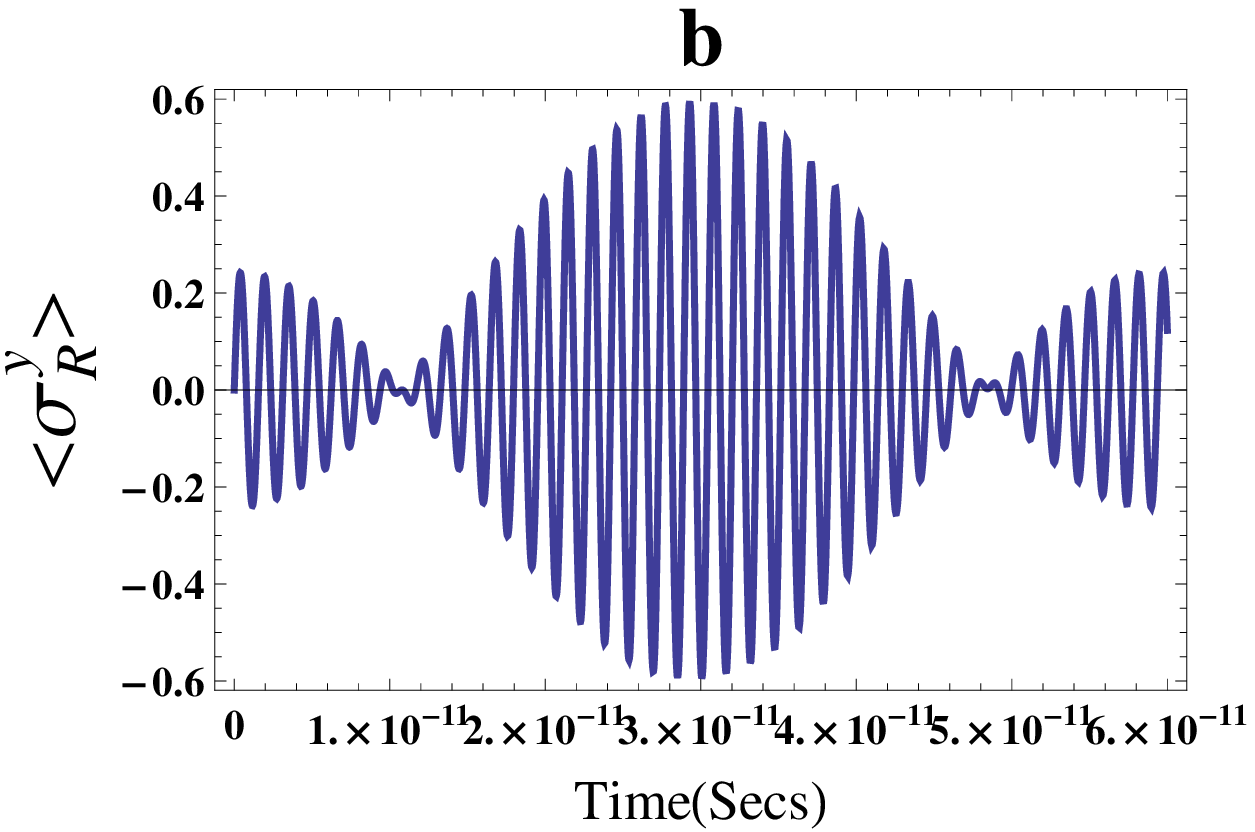}\\
\end{tabular}
\caption{Dynamics of $\sigma_{R}^{x}(t)$ (a) and  $\sigma_{R}^{y}(t)$ (b) for the non-identical width (NID) case for $B=6.4 T$. The various parameters used are: $\alpha=1 \times 10^{-9} eV-cm$, $\beta=0.3 \times eV-cm$, g=-0.45, $w_{x}=25 \sqrt{2} nm$, $w_{x}^{'}=1.0375 w_{x}$. }
\end{figure}\label{fig7}

\begin{figure}[t]
\hspace{-0.0cm}
\begin{tabular}{cc}
\includegraphics [scale=0.60]{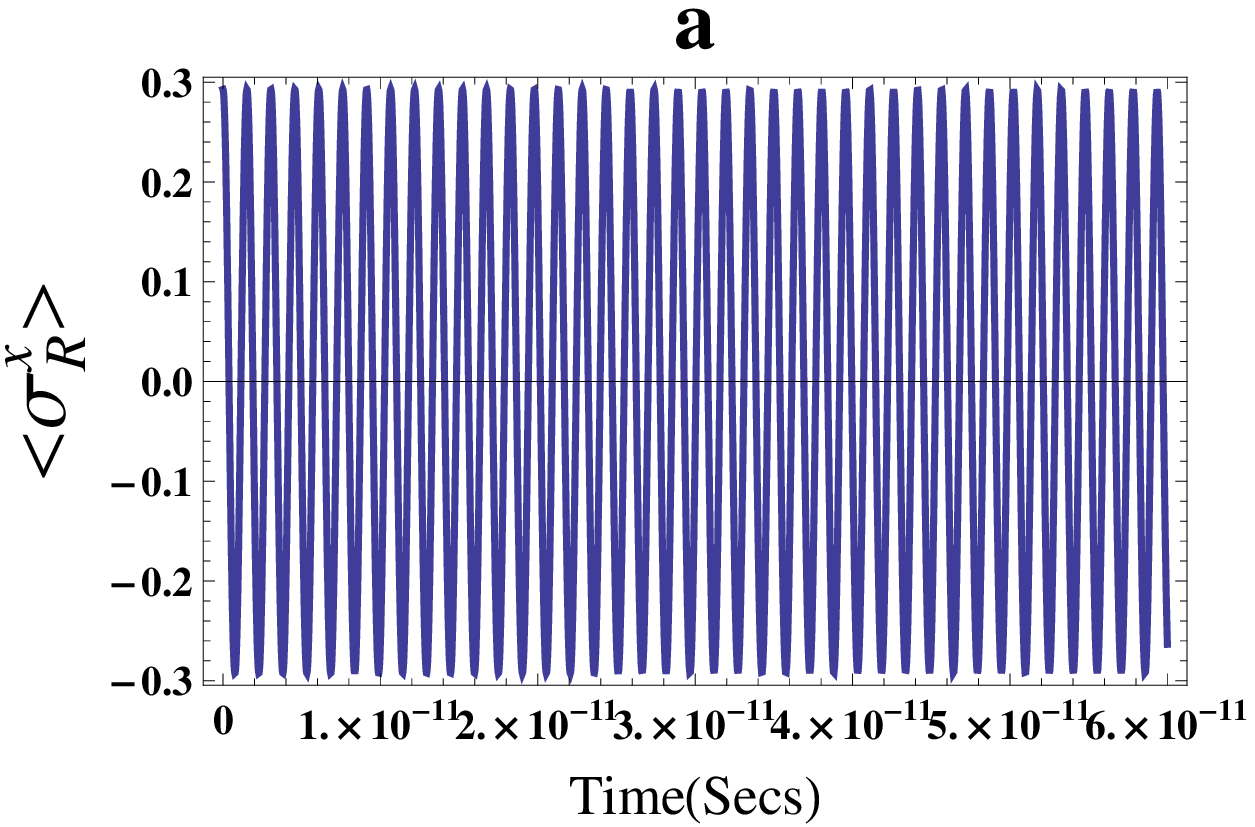}& \includegraphics [scale=0.60]{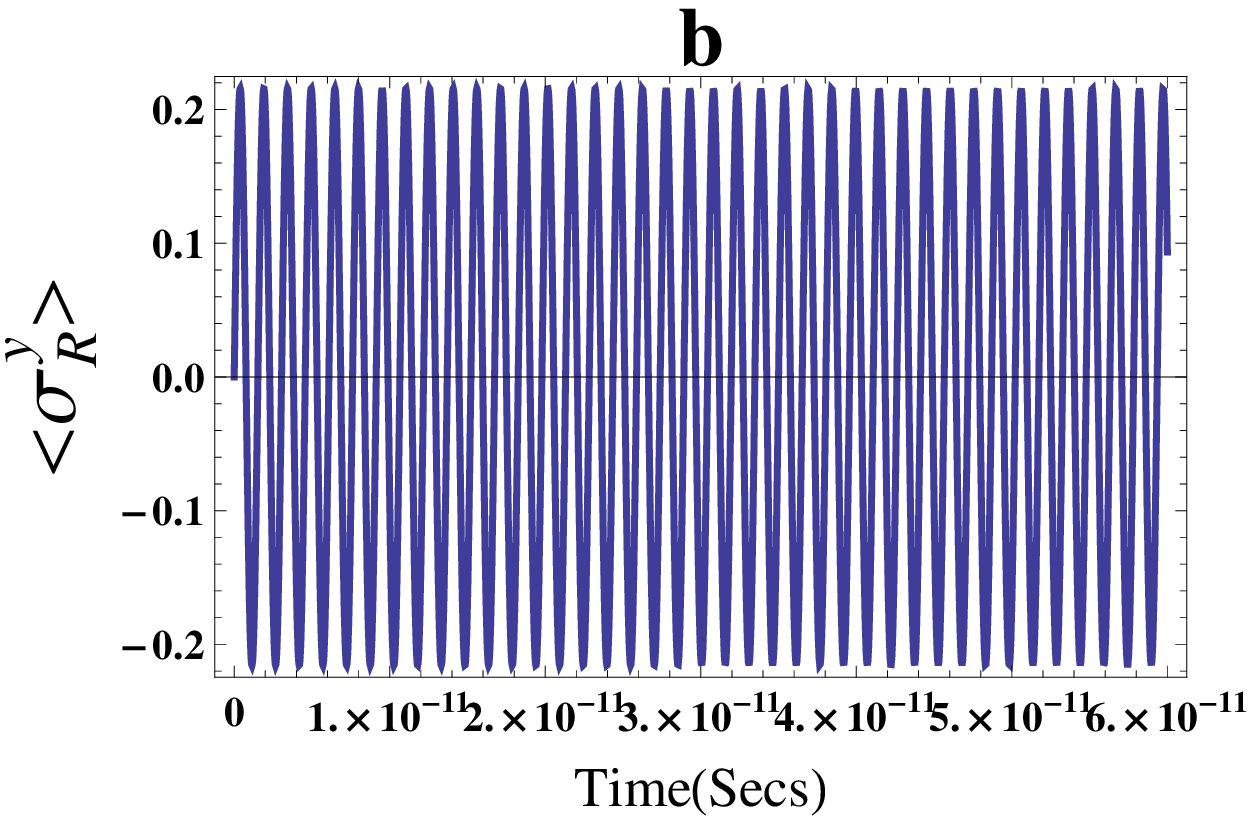}\\
\end{tabular}
\caption{Dynamics of $\sigma_{R}^{x}(t)$ (a) and  $\sigma_{R}^{y}(t)$ (b) for the non-identical width (NID) case for $B=11 T$. The various parameters used are: $\alpha=1 \times 10^{-9} eV-cm$, $\beta=0.3 \times eV-cm$, g=-0.45, $w_{x}=25 \sqrt{2} nm$, $w_{x}^{'}=1.0375 w_{x}$. }
\end{figure}\label{fig8}

\begin{figure}[t]
\hspace{-0.0cm}
\begin{tabular}{cc}
\includegraphics [scale=0.55]{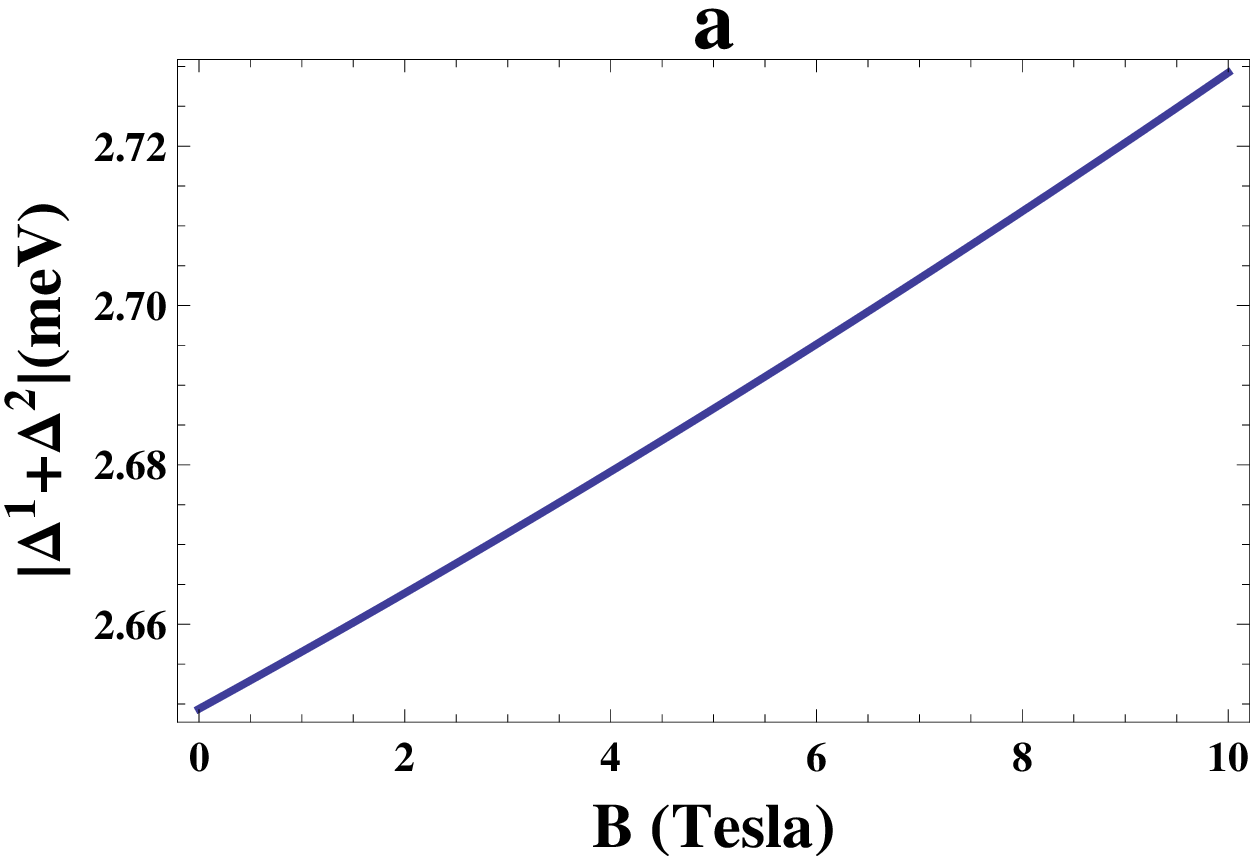}& \includegraphics [scale=0.55]{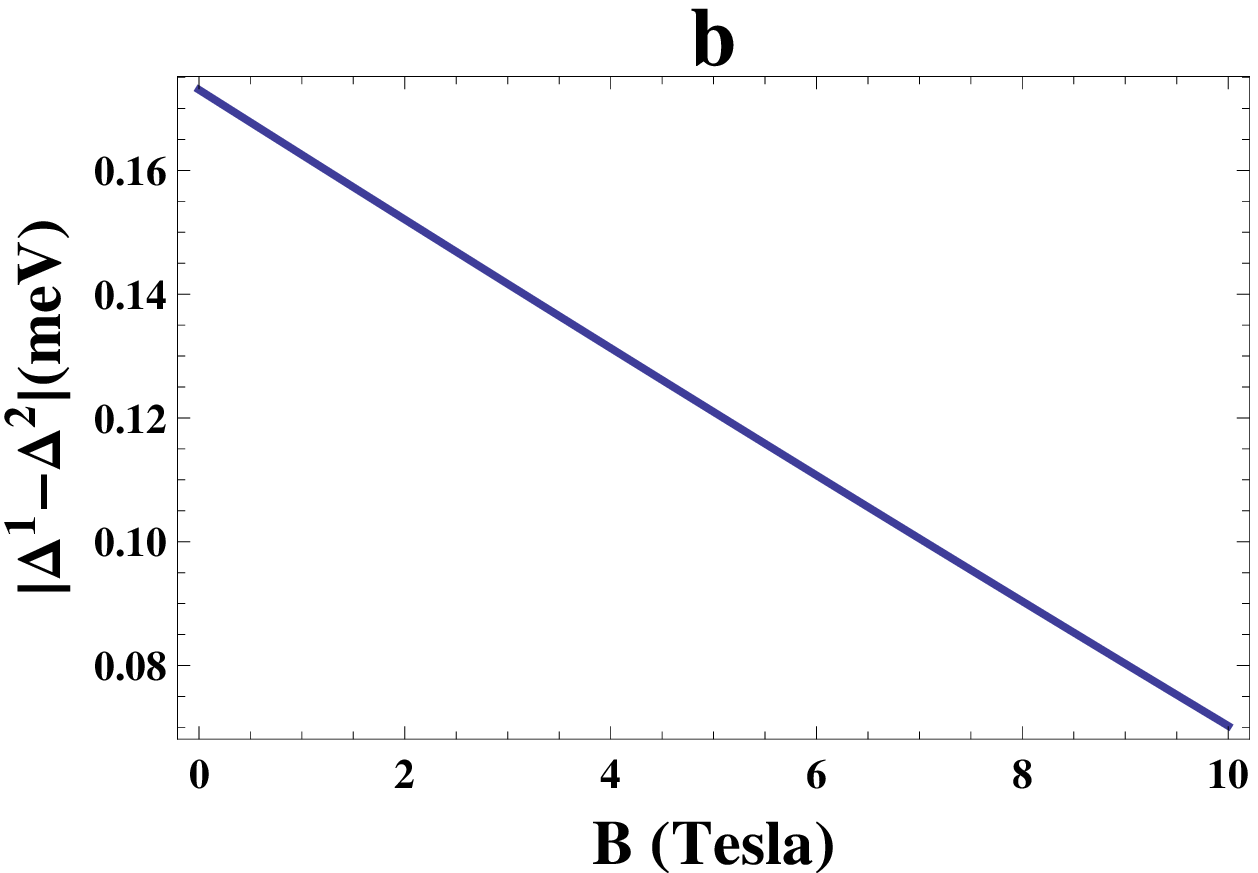}\\
\end{tabular}
\caption{ Plot of $\Delta^{(1)} + \Delta^{(2)}$ (a) and $\Delta^{(1)} - \Delta^{(2)}$ (b) as a function of the magnetic field .  }
\end{figure}\label{fig9}

\begin{figure}[t]
\hspace{-0.0cm}
\begin{tabular}{cc}
\includegraphics [scale=0.65]{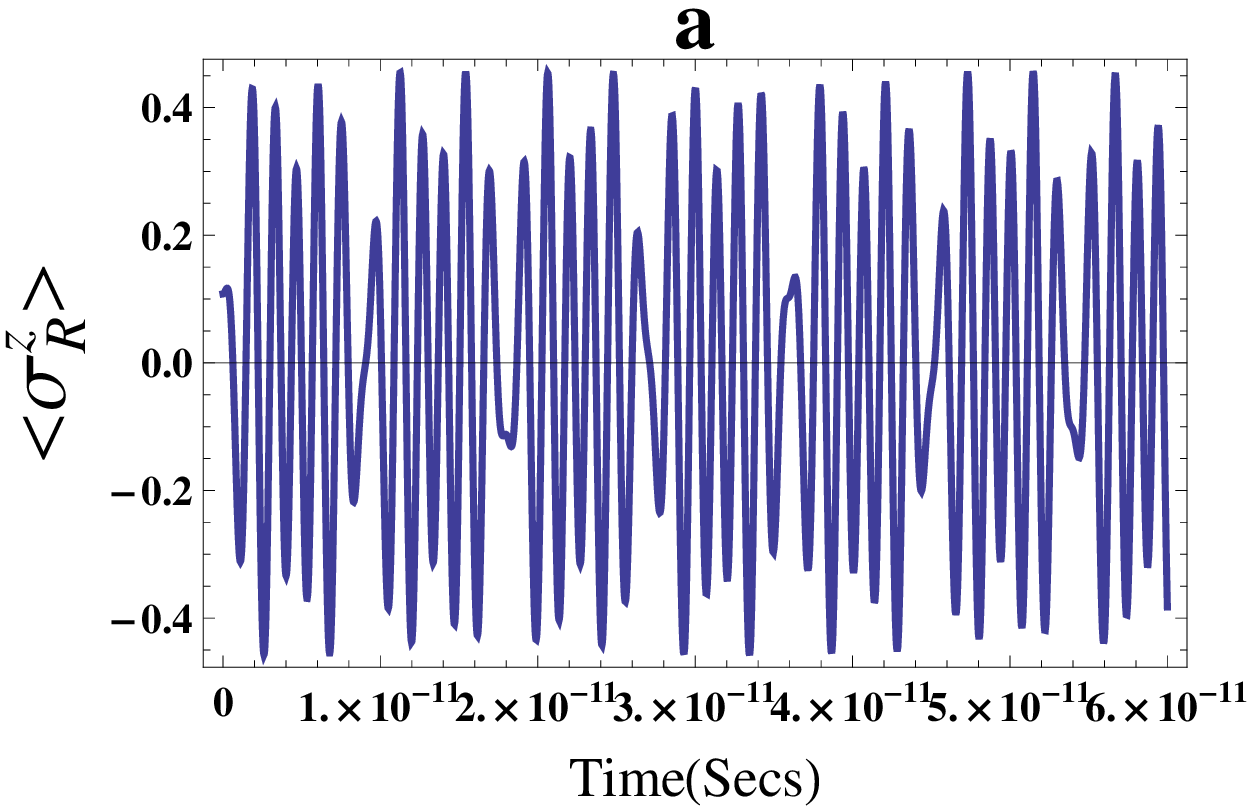}& \includegraphics [scale=0.60]{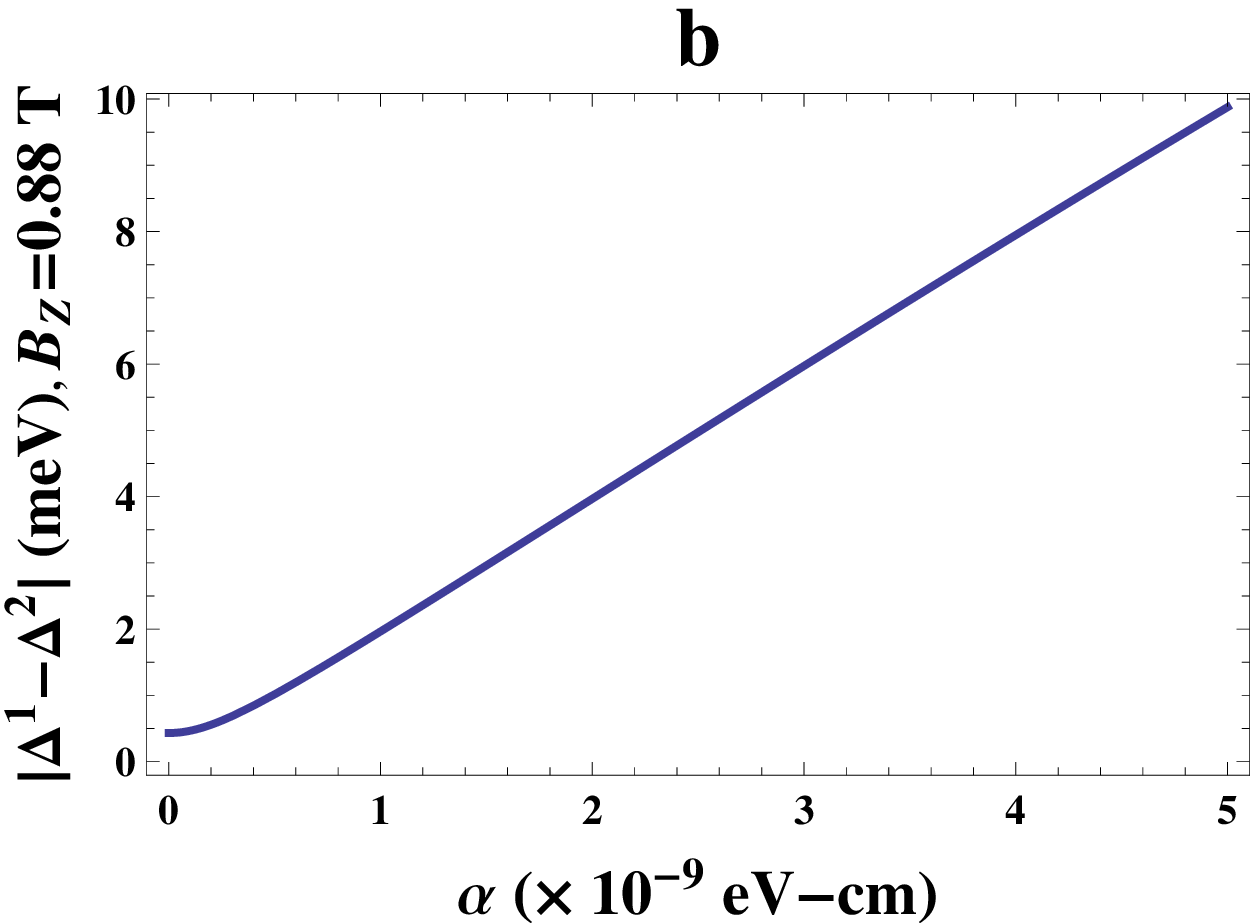}\\
\end{tabular}
\caption{Plot (a): Dynamics of $\sigma_{R}^{x}(t)$ (a) for the non-identical width (NID) case for $B=0.88 T$. The various parameters used are: $\alpha=2 \times 10^{-9} eV-cm$, $\beta=0.3 \times eV-cm$, g=-0.45, $w_{x}=25 \sqrt{2} nm$, $w_{x}^{'}=1.0375 w_{x}$. Plot (b): $\Delta^{(1)} - \Delta^{(2)}$ versus Rashba interaction strength ($\alpha$ ) for the NID case. }
\end{figure}\label{fig10}

\begin{equation}\label{7}
|\Psi>=\sum_{n}\xi_n(t) e^{\frac{-iE_nt}{\hbar}}|\Psi_n>,
\end{equation}

where $\xi_n(t)$ are the expansion coefficients . Once the wavefunctions are known, we can calculate the probability $ P_{R}(t)$ to find the electron in the right dot and expectation value of the $i^{th}$ spin component $\sigma_R^i$ in the right dot since we are interested in the quantum transport from the left to the right quantum dot.

\begin{equation}
P_R(t)=\int_{0}^{\infty} \Psi^{\dagger} \Psi dx,
\end{equation}

\begin{equation}
\sigma_{R} ^{i} (t)=\int_{0}^{\infty} \Psi^{\dagger} \sigma^{i}\Psi dx.
\end{equation}

The explicit expressions for $P_R(t)$ and $\sigma_R^i(t)$ are given in the appendix A.

\section{Result and Discussion}

We are interested in the inter dot transition. To demonstrate the nontrivial dynamics of the tunnelling and spin – flip process, we plot the dynamics of $P_R(t)$ and $\sigma_R^i(t) (i=x,y,z)$ for different magnetic fields.

For the ID case as shown in Fig.2 and Fig.3, the spin component $\sigma_R^z(t)$ displays an irregular pattern and the mean value of $\sigma_R^{z}(t)$ (denoted by $(\sigma_R^z(t))_{M}$) increases from $-0.10$ at $B_{z}=11T$ to $-0.25$ at  $B_{z}=0.88T$ .  Similarly the dynamics of  $P_R(t)$ is also irregular and  the mean value of $P_R(t)$(denoted by $(P_R(t))_M$) increases from $0.35$ at $B_z=11T$ to 0.52 at $B_z=0.88T$. The strong oscillations of  $P_R(t)$ around a mean value of $0.52$ for small and moderate fields $ B_z = 0.88T$ (Fig. 2b), $B_{z}=2.82T$ (not shown) and $B_{z}=6.4T$ (Fig. 3b), indicates that the external field almost equalizes the probabilities to find the electron in the right and left dot . For large magnetic field of $11T$ , the mean value around which $P_R(t)$ oscillates is $0.35$ (not shown) which shows that a strong magnetic field of $11 T$ suppresses the tunneling from the left to the right well. This indicates that for small magnetic fields, a small spin polarization can be achieved in the tunneling under ID case.

The dynamics for the NID case is more regular and rich as depicted in figure 4 and figure 5. The dynamics comprises of low frequency motion due to the term $\Delta^{(1)}-\Delta^{(2)}$ superimposed by high frequency oscillations due to the involvement of term $\Delta^{(1)}+\Delta^{(2)}$ in the expression for $P_R(t)$ and $\sigma_R^i(t)$ (see appendix).  For the NID case, $(\sigma_R^z(t))_M$ is zero for all magnetic fields while $(P_R)_M$ increases from $0.32$ at $B_z=11T$ to $0.54$ at $B_z=0.88T$, which indicates that no spin- polarization can be achieved in the tunneling.

The SO interaction is the sum of bulk originated Dresselhaus ($\beta$) and structure related Rashba ($\alpha$) terms. In the tunneling, spin precession around both the $x$ as well as $y$ axis is observed as shown in figure 6 and figure 7 for the NID case. The spin precession around the y axis occurs due to the Rashba SO term while the spin precession around the $x$ axis occurs due to the Dresselhaus SO term. The magnetic field produces the zeeman spin splitting of the level $\Delta_z =  E_{n\downarrow}-E_{n\uparrow}= \mid g\mid \mu_B B_z$ for the ID case. On the other hand for the NID case, the spin splitting energy is given by \citep{bhowmik},

\begin{equation}
\Delta_{z}= 2 \sqrt{ \left[ \frac{g \mu_{B} B}{2} \right]^2+ \alpha^2 \frac{64}{w_{x} w^{'}_{x}} f(w_{x},w^{'}_{x}) },
\end{equation}

where

\begin{equation}
f(w_{x},w^{'}_{x})= \cos^2(\frac{\pi w_{x}}{2 w^{'}_{x}}) \cos^2(\frac{\pi w^{'}_{x} }{2 w_{x} }) \left[ \frac{1}{((w_{x}/w^{'}_{x})^2-1)((w^{'}_{x}/w_{x})^2-1)} \right]^2.
\end{equation}

For the NID case, we note that the spin split energy is larger as compared to that for ID case. With the increase in magnetic field , the effect of SOC decreases, leading to smaller amplitudes of spin precession as can be seen by comparing figures 6 and 7 with figure 8. Comparing the dynamics of $<\sigma_{R}^{z}(t)>$ from Figs. 4 and Fig. 5, there is a minor increase in the Rabi frequency $|  \Delta^{(1)} + \Delta^{(2)}|$ with increasing magnetic field which is consistent with previous theoretical \citep{rashba1, rashba2} and experimental result \citep{nowack}. On the other hand the slow oscillations $|\Delta^{(1)} - \Delta^{(2)}|$ becomes slower with increasing magnetic field. These two observations are clearly depicted in Fig.9 where $\Delta^{(1)} + \Delta^{(2)}$ and $\Delta^{(1)} - \Delta^{(2)}$ are plotted as a function of the magnetic field in Fig.9a and Fig.9b respectively. The low frequency dynamics corresponds to the spectrum of the low energy states while the high frequency oscillations involve the higher-energy states. As evident from Figures 6,7 and 8, increasing the magnetic field, the slow oscillations of $<\sigma_{R}^{x}(t)>$ and $<\sigma_{R}^{y}(t)>$ gradually disappear at $B_{z}=11 T$. Thus at high magnetic fields, only the higher energy states are involved in the dynamics. The same conclusion cannot be reached for the ID case due to its irregular dynamics. In addition, we observe that for the ID case, both the initial state and spin precession axis change with magnetic field leading to different phase shifts between the spin components. On the other hand, for the NID case, the initial state and the spin precession axis remains intact for different magnetic fields. The incomplete Rabi spin flips results from the fact that the the electron tunneling between the potential minimas establishes a competing spin dynamics which prevents the electric field to flip the spin efficiently ($\sigma_R^z(t)$ to reach $\pm 1$). The slow tunneling dynamics of the NID case (Fig.4b and 5b) allows the electron to stay in the right well for a longer duration. The spin precession in this slow interminima motion induces a corresponding spin dynamics which does not allow the electric field to flip the spin efficiently and thus spin polarization is absent for the NID case. This is in contrast to the ID case where the tunneling dynamics is extremely fast and the electron oscillates rapidly between the left and the right well. The fast inter-minima motion washes out the induced spin dynamics and thus allows the electric field to flip the spin.

Spintronics based implementation of scalable quantum computers has generated much interest in exploring coherent control of sinle qubit rotation using Rabi oscillations \citep{bhowmik}. In order to successfully implement single qubit rotation selectively in a sample comprising of multiple quantum dots using Rabi oscillations, the spin splitting energy in the target quantum dot has to be changed appreciably using Rashba interaction \citep{bhowmik}. It was shown in ref. \citep{bhowmik} that the energies of the spin-split levels of the lowest subband in a QD decreases with increasing Rashba interaction strength. Since Rabi oscillations play an important role in the control of single qubit rotation, we have studied here the dynamics of $<\sigma_{R}^{z}(t)>$ as a function of Rashba interaction strength $\alpha$. Fig.10a shows the plot of $<\sigma_{R}^{z}(t)>$ as a function of time for $\alpha=2 \times 10^{-9} eV-cm$ and $B=0.88 T$. Clearly the frequency of the slow oscillations $| \Delta^{(1)} - \Delta^{(2)} |$ increases as compared to that in Fig. 4a where $\alpha=1 \times 10^{-9} eV-cm$. This is also depicted in Fig.10b where $| \Delta^{(1)} - \Delta^{(2)} |$ is seen to increase with $\alpha$. The influence of $\alpha$ on $|  \Delta^{(1)} + \Delta^{(2)}|$ is not appreciable. These observations indicates that increasing the Rashba interaction strength induces a rapid energy transfer between the low energy states while energy transfer between the higher energy states essentially remains uneffected.

\section{Conclusion}

In conclusion, we have studied the quantum spin and charge dynamics of a single electron confined in one-dimensional GaAs double quantum dot with rashba and Dresselhaus spin-orbit interaction. We have  made a comparitive study of two specific cases: In first case, the spatial parts of the wavefunctions are same for spin-up and spin-down states (ID case)  while in the second case we consider the spatial parts of the wavefunctions to be different for spin-up and spin-down states (NID case). We demonstrate that the dynamics of the NID case is more regular. The dynamics comprises of a slow as well as fast Rabi oscillations. The slow oscillations corresponds to dynamics involving low energy states while fast oscillations involve the higher energy states. For the ID case, we found a small amount of spin polarization in the tunneling but for the NID case, a complete absence of spin polarization in the tunneling is noticed. We found that a coherent control of the slow Rabi oscillations is possible using the Rashba  interaction strength and the external magnetic field which makes this scheme useful for single qubit manipuation for quantum computer applications.

\begin{acknowledgments}
A. B acknowledges financial support from the University Grants Commission, New Delhi under the UGC-Faculty Recharge Programme. M. S acknowledges financial support from the University Grants Commission for the junior research fellowship.
\end{acknowledgments}

\newpage

\section{Appendix A}

\textbf{Identical Width Case}

\begin{eqnarray}\label{A1}
\sigma_{R} ^ {y}(t)&=&N^{2}(0.84(a_{1} b_{2} - a_{2} b_{1})\sin(2\Delta ^{(1)} t) +0.998(d_{2} a_{1} -a_{2} d_{1}
+ b_{2} c_{1} - b_{1} c_{2}) \sin([\Delta ^{(1)}+\Delta ^{(2)}] t) \nonumber\\
&+& 0.998(c_{2} a_{1} -a_{2} c_{1} + b_{2} d_{1} - b_{1} d_{2}) \sin([\Delta ^{(1)}-\Delta ^{(2)}] t)
+0.84(c_{1} d_{2} - c_{2} d_{1}) \sin(2\Delta^{(2)} t);\nonumber\\
\sigma_{R} ^{z}(t)&=&N^{2}(\frac{1}{2}(a_{2} ^{2} - a_{1} ^{2} +b_{2} ^{2} - b_{1} ^{2} + c_{2} ^{2} - c_{1} ^{2}
+ d_{2} ^{2} - d_{1} ^{2})+(a_{2} b_{2} -a_{1} b_{1})\cos(2\Delta^{(1)} t) \nonumber\\
&+&(c_{2} d_{2} - c_{1} d_{1})\cos(2\Delta^{(2)} t) + \frac{8}{3\pi}(a_{2} c_{2} - a_{1} c_{1})\cos([\Delta^{(1)}-\Delta^{(2)}] t) \nonumber\\
&+&\frac{8}{3\pi}(a_{2} d_{2} - a_{1} d_{1} + b_{2} c_{2}- b_{1} c_{1})\cos([\Delta^{(1)}+\Delta^{(2)}] t));\nonumber\\
\sigma_{R} ^ {x}(t)&=&N^{2}(0.858(a_1 a_2+b_1 b_2 +c_1 c_{2} +d_1 d_2)+0.858(a_2 b_1 +a_1 b_2) \cos(2\Delta^{(1)} t) \nonumber\\
&+& 0.858(d_{1} c_{2} + c_{1} d_{2}) \cos(2\Delta^{(2)} t) + 0.998(a_{2} d_{1} +a_{1} d_{2} + b_{2} c_{1} +b_{1} c_{2})\cos([\Delta^{(1)}+\Delta^{(2)})] t) \nonumber\\
&+& 0.998(a_{2} c_{1}+a_{1} c_{2} +b_{2} d_{1} + b_{1} d_{2})\cos([\Delta^{(1)}-\Delta^{(2)}] t);\nonumber\\
P_{R}(t)&=&N^{2}(\frac{1}{2}(a_{2} ^{2} + a_{1} ^{2} +b_{2} ^{2} + b_{1} ^{2} + c_{2} ^{2} + c_{1} ^{2} + d_{2} ^{2} + d_{1} ^{2})
+\frac{8}{3\pi}(a_{2} d_{2} + a_{1} d_{1} + b_{2} c_{2} +b_{1} c_{1}) \cos([\Delta^{(1)} + \Delta^{(2)}] t) \nonumber\\
&+&\frac{8}{3\pi}(b_{2} d_{2} +b_{1} d_{1} +a_{1} c_{1} +a_{2} c_{2})\cos([\Delta^{(1)}-\Delta^{(2)}] t)
+(a_{1} b_{1}+a_{2} b_{2})\cos(2\Delta^{(1)}t) \nonumber\\
&+&(c_{1} d_{1} + c_{1} d_{2}) \cos(2\Delta^{(2)} t),
\end{eqnarray}

\newpage

\textbf{Non-identical width case}

\begin{eqnarray}\label{A1}
\sigma_{R}^{x}(t)&=&N^{2}(\frac{8}{3\pi}(a_{1} a_{2}+b_{1} b_{2}+c_{1} c_{2}+d_{1}d_{2})+\frac{8}{3\pi}(a_{2} b_{1} +a_{1} b_{2}) \cos(2\Delta^{(1)}t) \nonumber\\
&+&\frac{8}{3\pi}(c_{2} d_{1}+c_{1} d_{2}) \cos(2\Delta^{(2)}t) +(a_{2} c_{1} + a_{1} c_{2} +b_{2} d_{1}+b_{1} d_{2})\cos(\Delta^{(1)}t);\nonumber\\
\sigma_{R}^{y}(t)&=&N^{2}(\frac{8}{3\pi}(a_{1}b_{2}-a_{2}b_{1}\sin(2\Delta^{(1)}t)+\frac{8}{3\pi}(c_{1}d_{2}-c_{2}d_{1}\sin(2\Delta^{(2)}t) \nonumber\\
&+& (a_{1}c_{2}-a_{2}c_{1}+b_{2}d_{1}-b_{1}d_{2})\sin([\Delta^{(1)}-\Delta^{(2)}]t) +(a_{1}d_{2}-a_{2}d_{1}+b_{2}c_{1}-b_{1}c_{2})\sin([\Delta^{(1)}+\Delta^{(2)}]t);\nonumber\\
\sigma_{R}^{z}(t)&=&N^{2}(\frac{1}{2}(a_{2}^{2}-a_{1}^{2}+b_{2}^{2}-b_{1}^{2}+c_{2}^{2}-c_{1}^{2}+d_{2}^{2}-d_{1}^{2})
+(a_{2}b_{2}-a_{1}b_{1})\cos(2\Delta^{(1)} t) \nonumber\\
&+&(c_{2}d_{2}-c_{1}d_{1})\cos(2\Delta^{(2)} t) +\frac{8}{3\pi}(a_{2}d_{2}-a{1}d_{1}+b_{2}c_{2}-b_{1}c{1})\cos([\Delta^{(1)}+\Delta^{(2)}] t) \nonumber\\
&+& \frac{8}{3\pi}(a_{2}c_{2}-a_{1}c_{1})\cos([\Delta^{(1)}-\Delta^{(2)}] t);\nonumber\\
P_{R}(t)&=&N^{2}(\frac{1}{2}(a_{1}^{2}+a_{2}^{2}+b_{1}^{2}+b_{2}^{2}+c_{1}^{2}+c_{2}^{2}+d_{1}^{2}+d_{2}^{2})
+\frac{8}{3\pi}(a_{2}d_{2}+a_{1}d_{1}+b_{2}c_{2}+b_{1}c_{1})\cos([\Delta^{(1)}+\Delta^{(2)} t) \nonumber\\
&+&\frac{8}{3\pi}(b_{2}d_{2}+b_{1}d_{1}+a_{1}c_{1}+a_{2}c_{2})\cos([\Delta^{(1)}-\Delta^{(2)}] t)+(a_{1}b_{1}+a_{2}b_{2})\cos([2\Delta^{(1)}] t)\nonumber\\
&+&(c_{1}d_{1}+c_{2}d_{2})\cos(\Delta^{(2)})t).
\end{eqnarray}

\begin{eqnarray}
\mid a_{1}\mid^{2}&=&\frac{(-eE\eta+(\sqrt{\alpha^{2}+\beta^{2}})\sin(\frac{\pi w_{x}^{'}}{2 w_{x}})\frac{2\sqrt{w_{x} w_{x}^{'}}}{4w_{x}^{2}-(w_{x}^{'})^{2}})^{2}}{(\frac{\Delta E_{1}}{2}+\frac{\Delta_{z}}{2}+\Delta^{(1)})^{2}+(-eE\eta+(\sqrt{\alpha^{2}+\beta^{2}})\sin(\frac{\pi w_{x}^{'}}{2 w_{x}})\frac{2\sqrt{w_{x} w_{x}^{'}}}{4w_{x}^{2}-(w_{x}^{'})^{2}})^{2}};\\
\mid a_{2}\mid^{2}&=&1-\mid a_{1}\mid^{2};\\
\mid b_{1}\mid^{2}&=&\frac{(-eE\eta+(\sqrt{\alpha^{2}+\beta^{2}})\sin(\frac{\pi w_{x}^{'}}{2 w_{x}})\frac{2\sqrt{w_{x} w_{x}^{'}}}{4w_{x}^{2}-(w_{x}^{'})^{2}})^{2}}{(\frac{-\Delta E_{1}}{2}-\frac{\Delta_{z}}{2}+\Delta^{(1)})^{2}+(-eE\eta+(\sqrt{\alpha^{2}+\beta^{2}})\sin(\frac{\pi w_{x}^{'}}{2 w_{x}})\frac{2\sqrt{w_{x} w_{x}^{'}}}{4w_{x}^{2}-(w_{x}^{'})^{2}})^{2}};\\
\mid b_{2}\mid^{2}&=&1-\mid b_{1}\mid^{2};\\
\mid c_{1}\mid^{2}&=&\frac{(-eE\eta_{1}-(\sqrt{\alpha^{2}+\beta^{2}})\cos(\frac{\pi w_{x}^{'}}{w_{x}})\frac{2\sqrt{w_{x} w_{x}^{'}}}{w_{x}^{2}-4 (w_{x}^{'})^{2}})^{2}}{(\frac{\Delta E_{2}}{2}-\frac{\Delta_{z}}{2}-\Delta^{(2)})^{2}+(-eE\eta_{1}-(\sqrt{\alpha^{2}+\beta^{2}})\cos(\frac{\pi w_{x}^{'}}{w_{x}})\frac{2\sqrt{w_{x} w_{x}^{'}}}{w_{x}^{2}-4 (w_{x}^{'})^{2}})^{2}};\\
\mid c_{2}\mid^{2}&=&1-\mid c_{1}\mid^{2};\\
\mid d_{1}\mid^{2}&=&\frac{(-eE\eta_{1}-(\sqrt{\alpha^{2}+\beta^{2}})\cos(\frac{\pi w_{x}^{'}}{w_{x}})\frac{2\sqrt{w_{x} w_{x}^{'}}}{b^{2}-4 (w_{x}^{'})^{2}})^{2}}{(\frac{\Delta E_{2}}{2}-\frac{\Delta_{z}}{2}+\Delta^{(2)})^{2}+(-eE\eta_{1}-(\sqrt{\alpha^{2}+\beta^{2}})\cos(\frac{\pi w_{x}^{'}}{w_{x}})\frac{2\sqrt{w_{x} w_{x}^{'}}}{w_{x}^{2}-4 (w_{x}^{'})^{2}})^{2}};\\
\mid d_{2}\mid^{2}&=&1-\mid d_{1}\mid^{2};\\
\eta&=&\frac{-64w_{x} w_{x}^{'}\cos(\frac{\pi w_{x}^{'}}{w_{x}})+8\pi \sin(\frac{\pi w_{x}^{'}}{2 w_{x}}) (4w_{x}^{2}-(w_{x}^{'})^{2}}{\pi^{2} ((w_{x}^{'})^{2}-4w_{x}^{2})^{2} \frac{1}{(w_{x}w_{x}^{'})^\frac{3}{2}}};\\
\eta_{1}&=&\frac{64 w_{x} w_{x}^{'}\sin(\frac{\pi w_{x}^{'}}{2w_{x}})+16\pi \cos(\frac{\pi w_{x}^{'}}{w_{x}}) (w_{x}^{2}-4 (w_{x}^{'})^{2}}{\pi^{2} (-4 (w_{x}^{'})^{2}+w_{x}^{2})^{2} \frac{1}{(w_{x}w_{x}^{'})^\frac{3}{2}}};\\
N&=&\sqrt{\frac{1}{4+2(a_{1} b_{1}+a_{2} b_{2} + c_{2} d_{1}+ c_{2} d_{2})}}.
\end{eqnarray}

\begin{equation}
\Delta^{(1)}= \frac{1}{2} \sqrt{(\Delta E_{1}+\Delta_{z})^{2}+4 c^{2}},
\end{equation}

\begin{equation}
\Delta^{(2)}= \frac{1}{2} \sqrt{(\Delta E_{2}-\Delta_{z})^{2}+4 (c^{'})^{2}},
\end{equation}

\begin{equation}
c=-e E \eta+\sqrt{\alpha^{2}+\beta^{2}} \frac{2 \sqrt{b b'}}{4b^{2}-b'^{2}} \sin{\frac{\pi b'}{2 b}},
\end{equation}

\begin{equation}
c'=-e E \eta_{1}-\sqrt{\alpha^{2}+\beta^{2}} \frac{2 \sqrt{b b'}}{-4b^{2}+b'^{2}} \cos{\frac{\pi b'}{ b}},
\end{equation}

Here $\Delta E_{1}= E'_{2}-E_{1}$ and $\Delta E_{2}= E_{2}-E'_{1}$, where $E_{1}$ and $E_{2}$ are the lowest energies of $\uparrow$ state while $E'_{1}$ and $E'_{2}$ are the lowest energies of $\downarrow$ state. Here $a_{1},a_{2},b_{1},b_{2},c_{1},c_{2}, d_{1}$ and $d_{2}$ are the eigenvalues.


\begin{thebibliography}{1}

\bibitem{bird}
J. P. Bird, Electron transport in Quantum Dots,  (Spinger, Berlin, 2003).
%
\bibitem{kohler}
S. Kohler, J. Lehmann and P. Hanggi, Phys. Rep. \textbf{405}, 379 (2005).
%
\bibitem{recher}
P. Recher, E. V. Sukhorukov and D. Loss, Phys. Rev. Lett. \textbf{85}, 1962 (2000).
%
\bibitem{weinmann}
D. Weinmann, W. Hausler and B. Kramer, Phys. Rev. Lett., \textbf{74}, 984 (1995).
%
\bibitem{takahashi}
S. Takahashi and S. Maekawa, Phys. Rev. Lett., \textbf{80}, 1758 (1998).
%
\bibitem{liang}
W. Liang et al., Nature (London), \textbf{417}, 725 (2002).
%
\bibitem{park}
J. Park et al., Nature (London), \textbf{417}, 722 (2002).
%
\bibitem{merchant}
A. Merchant and N. Markovi \'{c}, Phys. Rev. Lett., \textbf{100}, 156601 (2008).
%
\bibitem{hamaya}
K. Hamaya et al., Phys. Rev. Lett., \textbf{102}, 236806 (2009).
%
\bibitem{burkard}
G. Burkard, D. Loss and D. P. DiVincenzo, Phys. Rev. B., \textbf{59}, 2070 (1999).
%
\bibitem{rashba1}
E. L. Rashba and AI. L. Efros, Phys. Rev. Lett. \textbf{91},  126405 (2003).
%
\bibitem{rashba2}
E. L. Rashba and AI. L. Efros, Appl. Phys. Lett. \textbf{83},  5295 (2003).
%
\bibitem{nowack}
K. C. Nowack , F. H. L. Koppens, Yu. V. Nazarov and L. M. K. Vandersypen, Science \textbf{318},  1430 (2007).
%
\bibitem{winkler}
R. Winkler, Spin-Orbit coupling effects in two-dimensional electron and hole systems, Springer tracts in modern physics, \textbf{191}, (Springer, Berlin, 2003).
%
\bibitem{pershinnd}
Y.V.Pershinnd C. Piermarocchi, Phys. Rev. B \textbf{72}, 245331 (2005).
%
\bibitem{sheng}
J. S. Sheng and K. Chang, Phys. Rev. B \textbf{74}, 235315 (2006).
%
\bibitem{rasanen}
E. Rasanen, A. Castro, J. Werschnik, A. Rubio, and E. K. U.   Phys.Rev. Lett. \textbf{98}, 157404 (2007).
%
\bibitem{kuan}
W.-H. Kuan, C.-S. Tang, and C.-H. Chang, Phys. Rev. B  \textbf{75}, 155326 (2007).
%
\bibitem{liu}
D.-Y. Liu and J.-B. Xia, J. Appl. Phys. \textbf{115}, 044313 (2014).
%
\bibitem{joibari}
F. K. Joibari, Ya. M. Blanter, and G. E. W. Bauer, Phys. Rev. B  \textbf{90} , 155301 (2014).
%
\bibitem{gumber}
S. Gumber,M. Gambhir, P. K. Jha, and Man Mohan J. Appl. Phys. \textbf{119}, 073101 (2016).
%
\bibitem{wolf}
S. A. Wolf et al., Science, \textbf{294}, 1488 (2001).
%
\bibitem{awschalom}
Semiconductor spintronics and quantum computation, editied by D. D. Awschalom, N. Samarth and D. Loss (Springer, Berlin, 2002).
%
\bibitem{petta}
J. R. Petta et al., Science, \textbf{309}, 2180 (2005).
%
\bibitem{sherman1}
D. V. Khomitsky, L. V. Gulyaev and E. Ya. Sherman Phys. Rev. B \textbf{85}, 125312 (2012).
%
\bibitem{sherman2}
D. V. Khomitsky and E. Ya. Sherman, Phys. Rev. B \textbf{79}, 245321 (2009).
%
\bibitem{sherman3}
D. Khomitsky and E. Sherman, Nanoscale Research Letters \textbf{6}, 212,(2011).
%
\bibitem{sherman4}
D. V. Khomitsky and E. Ya. Sherman, EPL, \textbf{90}, (2010) 27010.
%
\bibitem{bhowmik}
D. Bhowmik and S. Bandyopadhyay, Physica E, \textbf{41}, 587 (2009).
%


\end{thebibliography}
\end{document}